# Effect of ice sheet thickness on formation of the Hiawatha impact crater

Elizabeth A. Silber[*,1,2], Brandon C. Johnson[3,4], Evan Bjonnes[5], Joseph A. MacGregor[6], Nicolaj K. Larsen[7], Sean E. Wiggins[3]

[1]Department of Earth Sciences, Western University, London, ON, N6A 5B7, Canada
[2]The Institute for Earth and Space Exploration, Western University, London, ON, N6A 3K7, Canada
[3] Department of Earth, Atmospheric, and Planetary Sciences, Purdue University, West Lafayette, IN, USA
[4]Department of Physics and Astronomy, Purdue University, West Lafayette, IN, USA
[5]Department of Earth, Environmental and Planetary Sciences, Brown University, Providence, RI, USA
[6]Cryospheric Sciences Lab, NASA Goddard Space Flight Center, Greenbelt, MD, USA
[7]GLOBE Institute, University of Copenhagen, Copenhagen, Denmark

**Manuscript submitted for publication in EPSL**

*Corresponding author:

Elizabeth A. Silber

E-mail: esilber@uwo.ca

**Abstract**

The discovery of a large putative impact crater buried beneath Hiawatha Glacier along the margin of the northwestern Greenland Ice Sheet has reinvigorated interest into the nature of large impacts into thick ice masses. This circular structure is relatively shallow and exhibits a small central uplift, whereas a peak-ring morphology is expected. This discrepancy may be due to long-term and ongoing subglacial erosion but may also be explained by a relatively recent impact through the Greenland Ice Sheet, which is expected to alter the final crater morphology. Here we model crater formation using hydrocode simulations, varying pre-impact ice thickness and impactor composition over crystalline target rock. We find that an ice-sheet thickness of 1.5 or 2 km results in a crater morphology that is consistent with the present morphology of this structure. Further, an ice sheet that thick substantially inhibits ejection of rocky material, which might explain the absence of rocky ejecta in most existing Greenland deep ice cores if the impact occurred during the late Pleistocene. From the present morphology of the putative Hiawatha impact crater alone, we cannot distinguish between an older crater formed by a pre-Pleistocene impact into ice-free bedrock or a younger, Pleistocene impact into locally thick ice, but based on our modeling we conclude that latter scenario is possible.

## 1. Introduction

Recently, a putative impact crater with the diameter of $31.1 \pm 0.3$ km was discovered beneath the Hiawatha Glacier in northwestern Greenland (Fig. 1) (Kjær et al., 2018). The analysis of the glaciofluvial sediment samples collected from the river draining the structure shows the presence of shocked quartz, a marker indicative of meteoritic impact. Further, elevated concentrations of platinum-group elements (PGE) were found in the samples containing shocked quartz, and Kjær et al. (2018) further asserted that the putative impact crater may have been formed by a fairly rare iron asteroid. The size of the crater suggests that its formation likely caused significant regional – and perhaps even global – environmental perturbations (Toon et al., 1997; Erickson et al., 2020). As per scaling laws (Johnson et al., 2016b), to form a 31 km in diameter impact structure, an iron asteroid impacting at 17 km s$^{-1}$ at an incidence angle of 45º would have to be nearly 2 km wide (Collins et al., 2004). The probability of any composition asteroid of that size hitting Earth is low but non-negligible, occurring once every ~2 million years (Silber et al., 2018).

One of the major questions concerning the Hiawatha structure is its age. Although material suitable for radiometric dating has not yet been found and analyzed, radiostratigraphic and geomorphologic evidence suggest that the structure most likely was formed after the Pleistocene inception of the Greenland Ice Sheet (Kjær et al., 2018). This tentative conclusion was further supported by identification of impact-heated conifer wood fragments associated with impactite grains from the Hiawatha glaciofluvial outwash that is probably derived from the Early Pleistocene deposits (Garde et al., 2020). So, while the sum of available evidence is suggestive of a geologically young age, no firm evidence of its age yet exists.

Confirmed impact craters generally contain shock-diagnostic materials, such as extensive fracturing and brecciation, high-pressure minerals, and planar deformation features (PDFs) in quartz. Younger craters generally exhibit well-defined morphologic features and are less degraded than older impact structures (French and Koeberl, 2010; Melosh, 1989). For example, fresh craters feature relatively sharp and raised rims with overturned stratigraphy, and the lack of disrupted features (French and Koeberl, 2010; Melosh, 1989). Based on these identifiers, the putative Hiawatha impact structure might be relatively fresh. It has a rim-to-floor depth of $320 \pm 70$ m, and a dissected central uplift that is up to 50 m high and whose peaks are up to ~8 km apart (Kjær et al., 2018). For a subaerial impact (no ice present), simple modeling suggests that a fresh, 31-km-diameter subaerial crater would display a peak ring (Pike, 1985) and have a rim-to-floor depth of ~830 m (Collins et al., 2005). So, Hiawatha's morphology is muted compared to that expected for a subaerial impact, and yet it retains fundamental elements of a crater morphology (Fig. 1). One would also expect an impact of this size would blanket Greenland in rocky ejecta. If the impact occurred in the late Pleistocene, it is generally assumed that such ejecta should be easily identifiable within the six existing deep ice cores that typically record most of Last Glacial Period (115–11.7 ka) to the present day, including the Bølling-Allerød and the Younger Dryas (YD) transitions (Kjær et al., 2018). The YD is the millennium-long cold period that followed the Bølling-Allerød interstadial near the end of the last ice age at ~12.8 ka. Although there is a Pt anomaly of possible cosmic origin at the Bølling-Allerød/YD boundary in the Greenland Ice Sheet Project 2 (GISP2) ice core (Petaev et al., 2013), there is no other evidence of rocky ejecta in any ice cores (Seo et al., 2019) and substantial evidence challenging an impact that time (e.g., Sun et al., 2020). The presence of ice, however, would affect the morphology and depth of the final crater, as well as the distribution of rocky ejecta (Senft and Stewart, 2008).

Here we model several possible scenarios for the formation of the putative Hiawatha impact crater using the iSALE-2D shock physics code (Collins et al., 2004; Wünnemann et al., 2006). To understand the effect of a pre-impact ice sheet, we investigate how the presence of thick ice affects the crater morphology and the dynamics and placement of the distal ejecta blanket.

## 2. Methods

We model the formation of the putative Hiawatha impact crater using the iSALE-2D Eulerian shock physics code (Collins et al., 2004; Ivanov et al., 1997; Melosh et al., 1992; Wünnemann et al., 2006), which is based on the SALE (Simplified Arbitrary Lagrangian Eulerian) hydrocode solution algorithm (Amsden et al., 1980). This hydrocode has been used previously to model impacts on Earth and other planetary bodies, and its outputs compare well against laboratory experiments (e.g., Bray et al., 2014; Collins et al., 2002; Rae et al., 2019; Silber and Johnson, 2017).

Due to the model's axial symmetry, all impacts are assumed to be vertical, with the projectile striking surface at a velocity ($v$) of 12 km s$^{-1}$ (Collins et al., 2004). Because the most probable impact angle is 45°, the impact velocity we use represents the vertical component of the mean asteroidal impact velocity of 17 km s$^{-1}$ for Earth (Collins et al., 2004). The projectile diameter needed for an iron impactor to produce a 31-km-wide crater is approximately 1.8 km, which we adopt in this study (Johnson et al., 2016b).

The rocky target is assumed to be composed of granite (Pirajno et al., 2003), represented by the ANEOS-derived equation of state (EOS) for granite (Pierazzo et al., 1997). Terrestrial ice sheets are composed of ice Ih, represented by the Tillotson EOS (Ivanov et al., 2002; Tillotson, 1962). This approximation is consistent with earlier modeling studies of impacts on icy bodies (Bray et al., 2014; Cox and Bauer, 2015; Silber and Johnson, 2017). Following Kjær et al. (2018), the projectile is assumed to be metallic, represented by the ANEOS for iron (Thompson, 1990). The target surface temperature ($T$) was set to 250 K, and gravity ($g$) to 9.81 m s$^{-2}$. The thermal gradient (d$T$/d$z$) of the Earth's crust was set to 10 K km$^{-1}$. The temperature in the ice sheet is expected to be relatively uniform through much of the ice sheet and increase near the base of the ice sheet (e.g., Dahl-Jensen et al., 1998). For simplicity, here we assume a spatially uniform ice sheet temperature ($T = 250$ K), which is a sufficient approximation for the purpose of the problem investigated here. We also tested different uniform englacial temperatures of 240 and 260 K and these changes did not significantly affect our results.

Table 1 lists the strength and damage model parameters for the target (ice sheet, rock) and the projectile (iron). In our models, we included the effect of acoustic fluidization, a mechanism responsible for controlling the degree to which the target is weakened during the cratering process. The two model parameters describing the Block Model of acoustic fluidization are the decay time ($\gamma_\beta$) and the limiting viscosity ($\gamma_\eta$) of the fluidized target (Melosh, 1979), for which we used $\gamma_\beta = 300$ and $\gamma_\eta = 0.015$ (Table 1; Collins, 2014; Rae et al., 2019). We also implemented the dilatancy model in iSALE-2D using the parameters given by Collins (2014). Finally, the code also includes the implementation of viscoelastic-plastic ice rheology to account for any viscous contribution to material deformation (Johnson et al., 2016a).

We modeled impact scenarios with and without an ice sheet, aiming to evaluate the most likely conditions that existed at the time the putative Hiawatha impact crater formed. We varied the pre-impact thickness of the ice sheet ($t_{ice}$) from 0.5 km to 2 km, in increments of 0.5 km. The reasoning for implementing these scenarios is two-fold. First, we are interested in the degree to which the final crater morphology, such as the development of the central uplift, is affected by ice-sheet thickness, or by excluding the ice sheet altogether. Second, we assume that the presence of an ice sheet could inhibit the ejection of rocky material, and thus we investigate the placement of distal ejecta to obtain the thickness of this layer. In all our models, we used the parameters as described earlier in this section and given in Table 1. However, our two modeling targets (morphology and distal ejecta thickness) require a slightly different setup in terms of grid size, resolution and simulation period.

In Section 3.1, we focus on crater morphology using a grid resolution of 18 cells per projectile radius (CPPR), which corresponds to the cell size of 50 m. This resolution provides sufficient detail to resolve the final crater morphology while minimizing computational expense. For the sake of completeness, we also ran a suite of test simulations using a rocky asteroid to investigate the effect on crater morphology. Assuming the impactor to be of the same composition as the target rock and applying the scaling laws (Johnson et al., 2016b), the projectile diameter needed to obtain a 31 km wide crater is 2.4 km. To maintain the same cell size throughout all models, the CPPR in these simulations was set to 24.

In Section 3.2, we evaluate distal ejecta emplacement, because we can make a direct comparison with observations to constrain the conditions that may have existed when this impact structure formed. These simulations were performed at 200 CPPR, to optimize tracking of the material being ejected (e.g., Johnson and Melosh, 2014). Our simulations included the implementation of Lagrangian tracer particles allocated to track the location of a parcel of material. Using the velocity of ballistically ejected tracers, we calculate the tracer's ballistic trajectory and assume emplacement where the trajectory intersects the pre-impact surface. The thickness of ejecta is estimated by dividing the volume of ejecta in each 10-km-wide (radial) bin by the area of that bin.

Moreover, we ran a suite of simulations, also at 200 CPPR, to evaluate the ejecta emplacement in the case of a rocky asteroid impacting the surface.

In Section 3.3, we investigate the production of impact-induced melt. We ran the simulations at 50 CPPR for the iron impactor and 60 CPPR for the rocky impactor. To optimize the computation time, the simulations were ended after several seconds, as that is the sufficient length of time for the shock to propagate through the target and cause melting. Taking advantage of Lagrangian tracer particles, we track ice and rock separately and record the highest shock pressure ($P_{shock}$) these materials experience during the impact (Pierazzo et al., 1997). That information is then used to obtain the total volume of material shocked above the certain pressure threshold that is required to either partially or fully melt the given material. Following the work by Pierazzo et al. (1997), the peak pressures required to partially and fully melt and vaporize ice are as follows: 0.4 GPa (incipient melt), 3 GPa (total melt), 4.5 GPa (incipient vaporization), and 43 GPa (total vaporization). The peak shock pressures of 46 GPa and 56 GPa, respectively, are needed to partially and fully melt rock (granite) (Pierazzo et al., 1997). Finally, we also evaluate the volume of target rock that will be subjected to pressures between 10 GPa and 25 GPa, as PDFs form in this range (French and Koeberl, 2010).

## 3. Results and discussion

In this section, we describe the effect of ice thickness on crater morphology and distal rocky ejecta. The final crater diameter produced in all simulations is approximately 31 km, consistent with the putative Hiawatha impact structure.

### 3.1 Effect of ice thickness on crater morphology

Fig. 2 shows time series of the formation of an impact crater as a result of an iron projectile striking a 1.5-km-thick ice sheet overlying the rocky target. The time steps shown are $t$ = 5, 30, 75, 140 and 340 s. Upon impact, a tremendous amount of energy is released by the projectile into the target, sending a shockwave away from the point of origin. The expanding shockwave and following rarefaction wave set up an excavation flow, which opens a transient cavity (Fig. 2a,b). The earliest, fastest ejecta is always composed of near-surface material, in this case ice (Fig. 2a). The transient crater (Fig. 2b) subsequently collapses due to gravity. Note that at 250 K, ice is much weaker and deforms more readily than rock. As the crater collapses, a central uplift is produced and weak ice covered by rocky ejecta collapses into the crater (Fig. 2c). As the central uplift collapses, it pushes rocky material outward, which would normally produce a peak ring (Morgan et al., 2016). The outward collapsing rock material is met by inward collapsing weak ice (Fig. 2d) producing a complex ice–rock mixture where we would normally expect to find a peak ring (Fig. 2e). The inward collapse of weak ice pushes some rocky material towards the crater center, resulting in the formation of a rocky central peak and a final crater filled with ice (Fig. 2e). We note that iSALE hydrocode tracks the cratering process only until the final crater is formed (order of minutes), and it does not address either subsequent subglacial erosion or the longer-term thermal evolution of the impacted region after the crater is formed.

Fig. 3 shows the cross-sections of the final impact crater for all five scenarios modeled in this study: impact into the purely rocky material without the presence of ice (Fig. 3a), and impact into an ice sheet with thicknesses of 0.5, 1, 1.5, and 2 km (Fig. 3b-e, respectively). Our results show unambiguously that the morphology of the rocky portion of the resulting final crater is modulated by the presence and thickness of an ice sheet, a finding also consistent with previous studies (e.g., Senft and Stewart, 2008). While the final crater diameter in all models is the same, the overall appearance of the crater rim, the crater wall and the crater floor varies according to ice-sheet thickness (Fig. 3). The crater rim is most prominent if formed by an impact into rock; as ice thickness increases, the crater rim becomes less pronounced. This pattern is expected, because there is less rocky material available to form the rim, and more of the impact energy is expended into displacing ice. Modeled crater depth is measured from the rim to the deepest part of the crater interior to the disrupted peak ring. Without an ice sheet, the modeled crater depth is 1050 m. As ice thickness increases (0.5, 1, 1.5 and 2 km), crater depth generally decreases to 850, 550, 300 and 400 m, respectively. Our simulations with 1.5 or 2 km of pre-impact ice are roughly consistent with the present observed rim-to-floor depth of the putative Hiawatha (320 ± 70 m; Kjær et al., 2018).

In addition to their resulting crater depths, our simulations with 1.5 or 2 km of pre-impact ice thickness result in disrupted peak rings and central uplifts that are qualitatively consistent with the observed morphology of the putative Hiawatha impact crater. However, in simulations without ice and with an ice sheet up to 1 km thick, a peak-ring basin is still produced. Thus, the presence of a thicker ice sheet promotes the formation of central uplift and subdues the peak-ring that is otherwise expected at this size (e.g., Pike, 1985). In models with an ice sheet 1.5 to 2 km thick, the rocky portion of the final crater exhibits a central uplift, buried under ice (Fig. 3d,e). A disrupted peak ring may be more easily eroded than other parts of the crater. In the thinnest ice-sheet scenario (0.5 km), no ice overlies the final crater (Fig. 3b), and in the 1-km-thick ice-sheet scenario, the impact structure is only partially covered by ice that moved inward during crater collapse (Fig. 3c). Note that these models ignore any ice flow into the crater after the model period (minutes), but which is expected to occur from outside the impact-affected area. Lastly, if the impactor was composed of rocky material instead of iron, there is no substantial difference in the final crater morphology across all scenarios. This analysis indicates that the thickness of the ice sheet significantly influences the morphological expression of the resulting impact structure.

We note the recent discovery of a possible second – and slightly larger – impact crater beneath the northwestern Greenland Ice Sheet (MacGregor et al., 2019). This second structure's diameter and depth are estimated at 36.5 km and 160 ± 100 m, respectively, so it is more degraded and likely older than the putative Hiawatha impact crater. However, it appears to possess a more dispersed and degraded central uplift than the putative Hiawatha impact crater, making it closer to a nascent peak-ring morphology. Our simulations suggest tentatively that the potential peak-ring morphology of this putative crater is more consistent with formation before the inception of the Greenland Ice Sheet.

### 3.2 Effect of ice thickness on distal rocky ejecta

Early in the cratering process, near-surface material is ejected at high velocity and the earliest ejecta has the highest velocities (Melosh, 1989, Johnson and Melosh, 2014). The behavior of the ejecta curtain when the ice sheet is present is best illustrated in Fig. 2a at $t = 5$ s. At this time, the earliest fastest ejecta are composed only of ice. Another factor limiting ejection of rocky material is the large contrast in target properties, with ice and rock responding to shock loading in a different manner. This assertion is consistent with previous studies that examined the dynamics of ejecta for impacts into icy layers (Senft and Stewart, 2008). Therefore, we should not expect to find rock-dominated ejecta far from the impact point.

The thicknesses of distal (>200 km) rocky ejecta as a function of radial distance from the crater center for an iron asteroid are shown in Fig. 4a and a rocky asteroid in Fig. 4b, with a particular focus on the thickness of ejecta at existing deep ice-core sites in Greenland. In Fig. 5, we include the map of modern Greenland with each panel showing modeled thickness of rocky ejecta produced by an iron impactor for all five scenarios (no ice, and ice thickness of 0.5, 1, 1.5 and 2 km).

As radial distance from the crater center increases, ejecta thickness decreases. In addition to reducing the thickness of rocky ejecta, the ice sheet limits the distance that rocky ejecta travel (Fig. 4). For an iron asteroid, the maximum distance that any significant quantity of rocky ejecta (≥0.01 mm) travel are 691, 636, 479, and 245 km for pre-impact ice-sheet thicknesses of 0.5, 1, 1.5, and 2 km, respectively. For a rocky asteroid, these distances are 690, 440, 420, and 320 km for pre-impact ice-sheet thicknesses of 0.5, 1, 1.5, and 2 km, respectively. Note that the larger rocky impactor produces more distal rocky ejecta than the smaller iron impactor. Beyond the above distances, all ejecta are composed of ice only. In our simulations there is no impactor material in the fast ejecta. As we discuss later, impactor material will be ejected downrange during in oblique impacts and may also be incorporated in the vapor plume. It is also possible that small amounts of impactor are incorporated in jetted material (Johnson et al., 2014).

Based on this modeling, we next re-evaluate the significance of the present lack of evidence of any impact ejecta in Greenlandic deep ice cores. The deep ice core sites are located at distances ranging from 210 to 1030 km away from Hiawatha Glacier, and reach depths from just over 1 km to approximately 3 km, recording ice as old as the Eemian period (130–115 ka; Dahl-Jensen et al., 2013) in the readily interpretable meteoric ice and yet older ages (approaching 1 Ma) in the more complex basal ice in central Greenland (Yau et al., 2016). The ice-core distances and core thicknesses are: Camp Century (210 km, ~1.2 km) NEEM (378 km, ~2.5 km), NorthGRIP (721 km, ~3.1 km), GISP2 (1011 km, ~2.8 km), and GRIP (1030 km, ~1.8 km). Note that DYE-3 is not included in this analysis due to its greater distance (1673 km) from Hiawatha Glacier. In all scenarios with pre-impact ice cover, between 0.1 and 10 mm of rocky ejecta is expected at the position of the closest deep ice core, Camp Century. However, that ice core is also the oldest of the six ice cores and less intensely studied than the later, more distal ice cores (Dansgaard et al., 1969; Johnsen et al., 1972; Kjær et al., 2018). At NEEM, the most recent ice core and 378 km away, less than 0.1 mm of rocky ejecta is expected for pre-impact ice-cover scenarios of 1.5 or 2 km. For all other ice cores, only 1 km of pre-impact ice cover is needed to result in no or negligible rocky ejecta being deposited at those sites. Note that this assessment ignores the effect of horizontal ice flow, which leads to the sourcing of each ice core's present ice column from up to tens of kilometers farther upstream, closer to the central ice sheet (e.g., Dahl-Jensen et al., 2003). This analysis suggests that – if the Hiawatha impact occurred during the Last Glacial Period – its ejecta is unlikely to be well preserved at most existing Greenland ice cores except for Camp Century, where it would likely have been detected during conventional core logging if its thickness was ~1 mm or greater.

We also generated a rough estimate on the proximal (<200 km) rocky ejecta blanket thickness (Fig. 4c,d). Because it is computationally prohibitive to run simulations at a very high resolution (200 CPPR), we used the 'normal' resolution outputs to generate these results. The inward flow of ice after ejecta emplacement makes a thickness estimate within 100 km of the crater less useful.

For several reasons, our model estimates of the thickness of rocky ejecta from the target should be considered as upper limits. First, oblique impacts into ice sheets are expected to further limit ejection of rocky material (Stickle and Schultz, 2012). Additionally, a continuous ejecta blanket is not expected to occur at distances greater than about one crater diameter from the crater rim (Melosh, 1989). Beyond that distance, ejecta are expected to be thinner and patchier. A related effect of oblique impacts is the wedge of avoidance, i.e., a wedge-shaped region uprange of the impact where ejecta is generally absent (e.g., Ekholm and Melosh, 2001). The arc of the wedge of avoidance is related to the impact angle, i.e., as the impact angle decreases, the wedge of avoidance increases. For example, a projectile entering the obliquely at 20° and 40° produces the wedge of avoidance with the angular size of 45° and 115°, respectively (Ekholm and Melosh, 2001). Therefore, depending on the angle and direction of impact, the wedge of avoidance could be so large that the ejecta would be absent at the ice-core sites for a yet-thinner ice sheet than that implied by our vertical-impact modeling.

Another possible source of rocky ejecta is the impactor, depending on its composition. During an oblique impact much of the impactor material will be deposited downrange of the impact (e.g., Pierazzo and Melosh, 2000a). Even for a vertical impact, an impact vapor plume may produce distal ejecta layers (e.g., Johnson and Melosh, 2012). Accurate models of the distribution of impactor material would require full 3D simulation with updated EOS that more accurately account for silicate vaporization (Kraus et al., 2012; Kurosawa et al., 2012). Without knowledge of the impact direction and full 3D simulation of the impact we cannot estimate the possible contribution of impactor material. However, to obtain an order-of-magnitude upper limit on how thick the impactor-originating ejecta could be, we assume that the entire rocky impactor was distributed as a uniform thickness layer with radial extent equal to the location of the various drill cores. Under this assumption this layer could be 25.5, 8, 2.2, 1.1, and 1.1 mm for core distances of 210, 378, 721, 1011, and 1030 km, respectively.

Considering that ice dominates the fast – and therefore distally emplaced – ejecta, the apparent absence of rocky ejecta in existing ice cores does not in and of itself rule out the possibility that the putative Hiawatha impact crater formed during the time period spanned by the ice cores, i.e., most of the Last Glacial Period. In other words, based on our modeling, a possible Late Pleistocene timing for the putative Hiawatha impact crater formation cannot be discounted if it occurred through a sufficiently thick ice sheet. This result is consistent with the preliminary age constraints presented by Kjær et al. (2018). The lack of ejecta in ice cores does not necessarily rule out a Last Glacial Period age, including the YD period, for the impact if the ice sheet was at least 1.5 km thick there at the time of the impact. However, we note that the ICE-6G model does not predict ice thicker than ~1 km at Hiawatha Glacier for the past 26 kyr (Stuhne and Peltier, 2015). Our modeling suggests that further investigation of the Camp Century ice core for ejecta or impactor signatures could more robustly rule out a Late Pleistocene age for the putative Hiawatha impact crater.

**3.3 Impact-induced melting**

In Table 2, we summarize our results by outlining the peak shock pressures and the impact-induced melt volumes for both an iron asteroid and a rocky asteroid, separated out by the target material (ice or rock). Fig. 6 shows the provenance plot of peak shock pressures reached within the material upon the impact by a 1.8 km wide iron asteroid (Fig. 6a-c), and a 2.4 km wide rocky asteroid (Fig. 6d-f). The iron asteroid produces greater shock pressures overall compared to its rocky counterpart, which is expected because of its higher density. In both cases, the peak shock pressures are more than sufficient to melt or vaporize ice and melt rock, but a sufficiently thick ice sheet also dissipates shock propagation.

Unless the impact is very shallow (<30º), axisymmetric models can still provide good estimates for melt production in oblique impacts (Pierazzo and Melosh, 2000b). If the ice sheet was 1.5–2 km thick at the time the putative Hiawatha crater formed, the impact by an iron asteroid would have melted 106–164 km$^3$ of ice, comparable to the amount of water in Lake Tahoe, USA. A rocky extraterrestrial body would produce an even greater volume of melted ice, 141–217 km$^3$ (Fig. 7a). Moreover, about 4 km$^3$ of ice would completely vaporize (Table 2). It should be noted that our simulations account for impact-induced melting only (i.e., immediately upon the impact), which means that any post-impact ice melting or water retention due to the presence of the melt sheet in the rocky target is not accounted for. Therefore, our ice melt estimates represent a lower limit. Future studies should consider the effect of melt sheet on post-impact evolution of the region, melting of the inflowing ice sheet, and the resulting hydrothermal activity.

In terms of rocky target material within a melt sheet produced with an iron impactor, ~32–36 km$^3$ of rock would be fully melted, regardless of ice thickness (Fig. 7b). In the case of a rocky asteroid, the rocky melt volume would be lower, but still substantial (~19–28 km$^3$; Fig. 7b). Finally, we determined the volume of rocky target material experiencing peak shock pressures conducive to the formation of PDFs (Fig. 7c). Up to 990 km$^3$ (iron asteroid) and 750 km$^3$ (rocky asteroid) of target rock could potentially be exposed to $P_{shock} > 10$ GPa.

## 4. Conclusions

The recent discovery of the putative 31-km-wide Hiawatha impact crater beneath the Greenland Ice Sheet reinvigorated interest in ice-affected impact processes (Kjær et al., 2018). We used iSALE-2D shock physics code to model possible formation scenarios for this putative impact crater and investigate the resulting morphology and the emplacement of distal rocky ejecta to infer possible conditions at the time of crater formation. The morphology of the simulated crater is qualitatively consistent with present observations if the ice sheet is 1.5–2 km thick, implying that the crater could have formed geologically recently if thick ice were present there at the time of impact (e.g., during a Pleistocene stadial). We also find that the presence of an ice sheet inhibits ejection of rocky material and that no rocky ejecta should be expected at distances exceeding 245 km for a 2-km-thick ice sheet. Thus, ignoring subsequent erosion, our results are consistent with the existing hypothesis that the putative Hiawatha impact crater formed after inception of the Greenland Ice Sheet around 2.6 Ma (Bierman et al., 2016). Further, its possible formation during the Last Glacial Period or at the onset of YD cannot yet clearly be ruled out based on the lack of rocky ejecta in existing ice cores alone. If the Hiawatha structure is confirmed and predates inception of the ice sheet, it's rim and porous peak ring (Rae et al., 2019) must have been preferentially eroded at relatively low rates to explain the observed morphology. While radiometric dating of this putative crater remains a priority for understanding when and how it formed, our study demonstrates the value of numerical modeling for contextualizing the history of impacts into ice sheets on Earth and elsewhere in the Solar System.

## Acknowledgements

We gratefully acknowledge the developers of iSALE-2D (www.isale-code.de), the simulation code used in this work, including Gareth Collins, Kai Wünnermann, Dirk Elbeshausen, Boris Ivanov and Jay Melosh. Some plots in this work were created with the pySALEPlot tool written by Tom Davison. All data associated with this study are listed in tables in the supporting information and shown in figures. The simulations were performed using iSALE-2D, version Dellen r-2114. The simulation inputs and model outputs are available on Harvard Dataverse (https://doi.org/10.7910/DVN/X6KZJG). We thank K. H. Kjær (GLOBE Institute, University of Copenhagen) for valuable discussions. NKL thanks the Carlsberg Foundation, Aarhus University

Research Foundation and the Villum Foundation for supporting this study. EB thanks the Bevan and Mary French Fund for Meteorite Impact Geology for partly supporting this research. The authors also thank Bill McKinnon and the two anonymous reviewers for their comments that helped improve our paper.

**Competing interests:** The authors declare no competing interests.

**List of Figures**

**Figure 1:** Bed topography beneath and in the vicinity of Hiawatha Glacier, northwestern Greenland (Kjær et al., 2018) overlain on hillshaded surface elevation (10-m ArcticDEM (Digital Elevation Model) mosaic; Porter et al., 2018). Symbology follows radar-identified features described by Kjær et al. (2018). Ice margin is from the Greenland Ice Mapping Project (Howat et al., 2014).

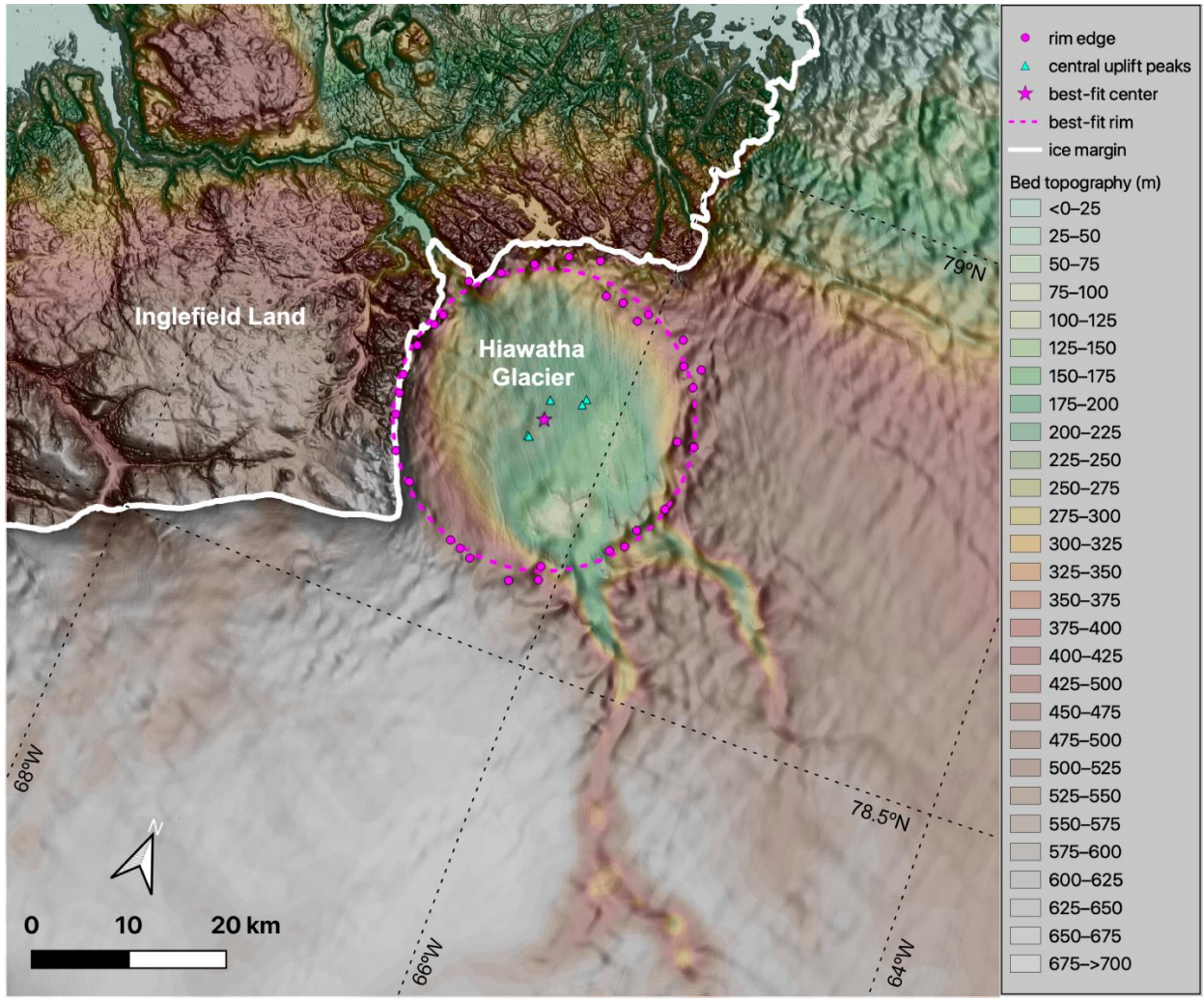

**Figure 2:** Time series of a modeled impact into a 1.5-km-thick ice sheet. Material is colored according to material type; dark brown, light brown and blue represent the iron impactor, granitic crust and ice, respectively. Axis origin marks the point of impact. Originally vertical and horizontal gray lines connect Lagrangian tracers and track deformation as the impact progresses. Dark brown blobs in the center represent the remnants of the iron impactor; however, we note that details along the vertical axis in 2D axisymmetric calculations are less reliable.

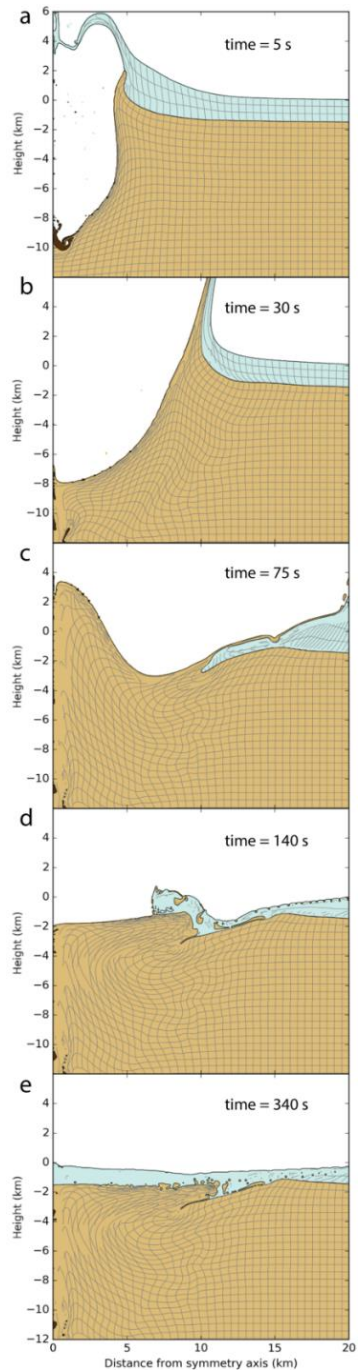

**Figure 3:** The cross-section of the final crater formed as a result of an impact into (a) a purely rocky target (no ice) and (b-e) an ice sheet with varying thickness. In the panels (b-e), top to bottom, the ice sheet thicknesses are 0.5, 1, 1.5 and 2 km. Colors and grid scheme follow Fig. 2.

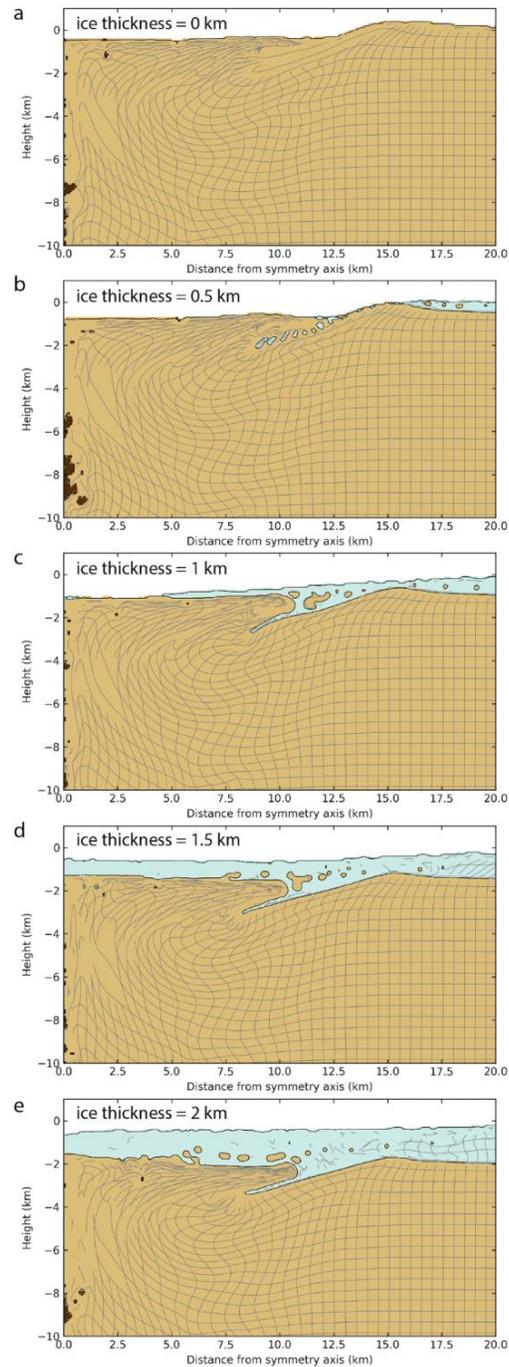

**Figure 4:** Thickness of distal rocky ejecta as a function of radial distance from the point of impact for (a) an iron and (b) a rocky asteroid. Runs are at 4.5-m resolution (200 CPPR) with pre-impact ice thickness indicated in the legend. Vertical lines mark the distance of ice cores (DYE-3 is 1673 km away and not included here). The maximum distance that any rocky ejecta travel are 691, 636, 479, and 245 km (iron asteroid) and 518, 384, 241, and 276 km (rocky asteroid) for pre-impact ice thickness of 0.5, 1, 1.5, and 2 km, respectively. Also is shown thickness of proximal ejecta extending up to 200 km for (c) and iron and (d) a rocky asteroid. Note that these runs are at 50-m resolution; while these do not capture ejecta in as great detail as high resolution runs, they offer a reasonable approximation.

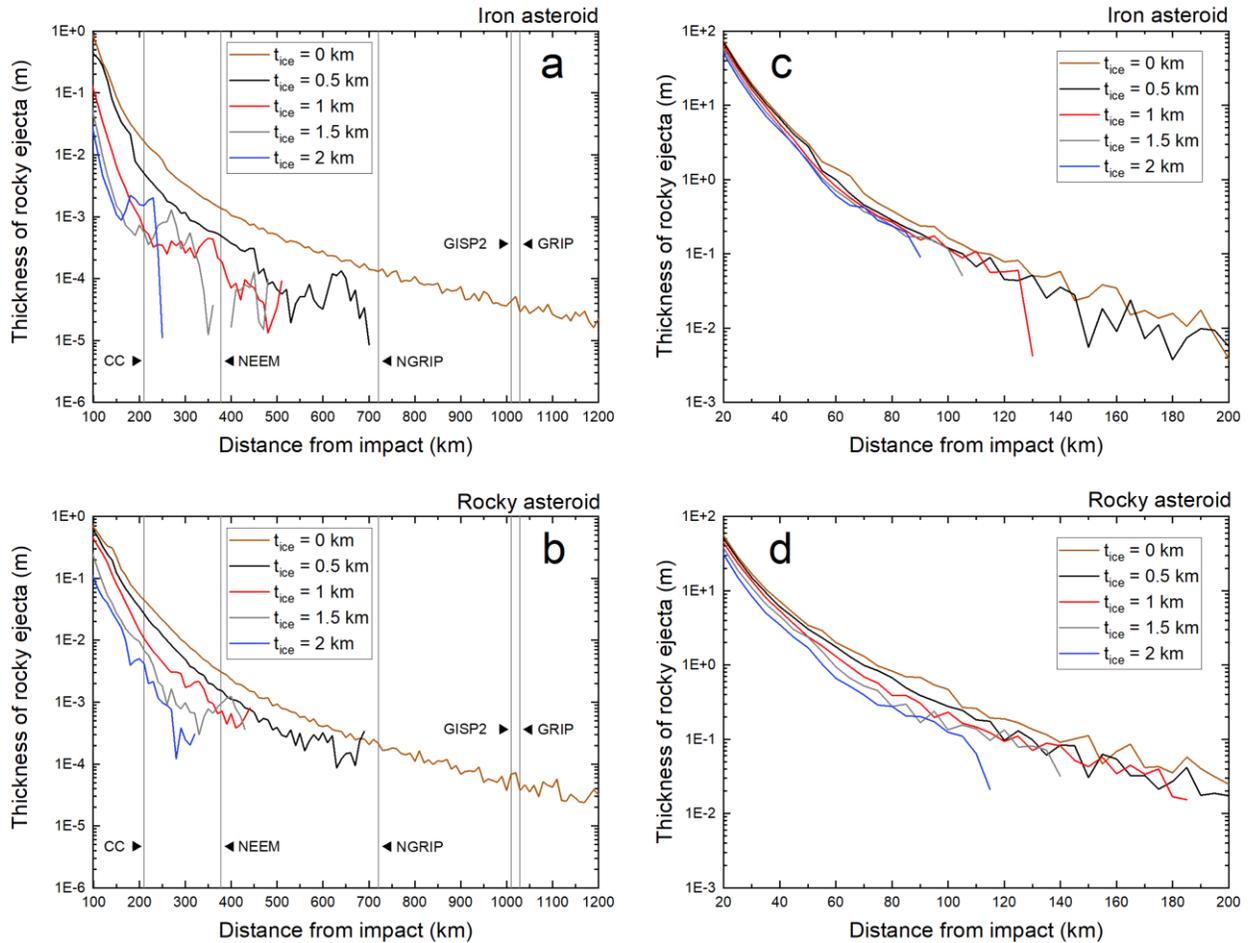

**Figure 5:** Map of modern Greenland with each panel showing modeled thickness of rocky ejecta due to impact of an iron asteroid, assuming no ice present ("no ice sheet") and then for all four considered Hiawatha ice-target scenarios (0.5, 1, 1.5 and 2 km, respectively).

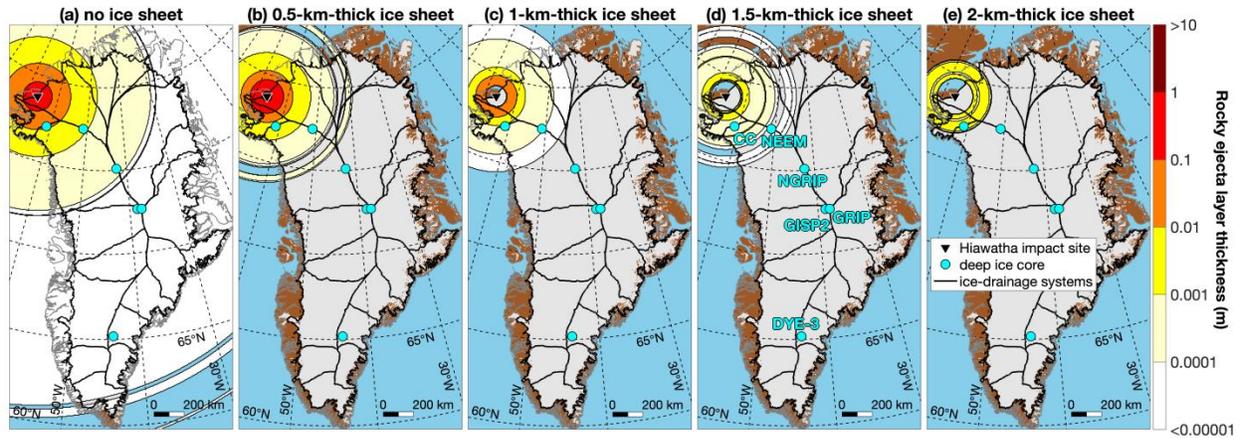

**Figure 6:** Provenance plot of peak shock pressures reached within the material 0.5 s and 0.7 s after impact by a 1.8 km wide iron asteroid (a-c), and a 2.4 km wide rocky asteroid (d-f), respectively. For easier visualization, the color bars represent the same scale across all panels (0–200 GPa). While the iron asteroid produces overall greater shock pressures than its rocky counterpart, in both cases the shock pressures are sufficiently high to readily melt ice and rock, while the ice sheet somewhat dissipates shock propagation.

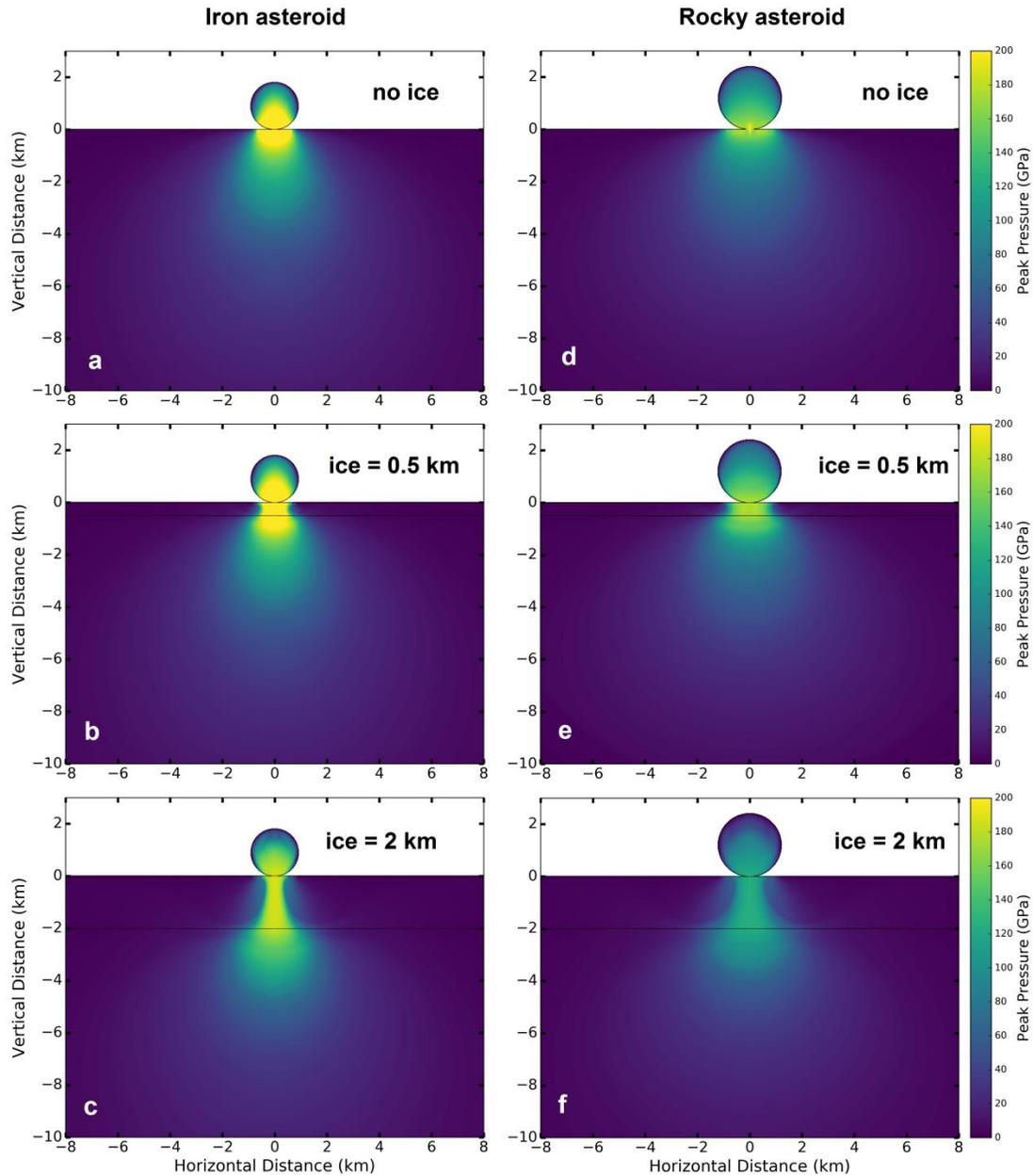

**Figure 7:** (a) Volumes of partially and fully melted ice; (b) volumes of partially and fully melted rock; (c) the volume of rock subjected to pressure range conducive to PDF formation (10–25 GPa).

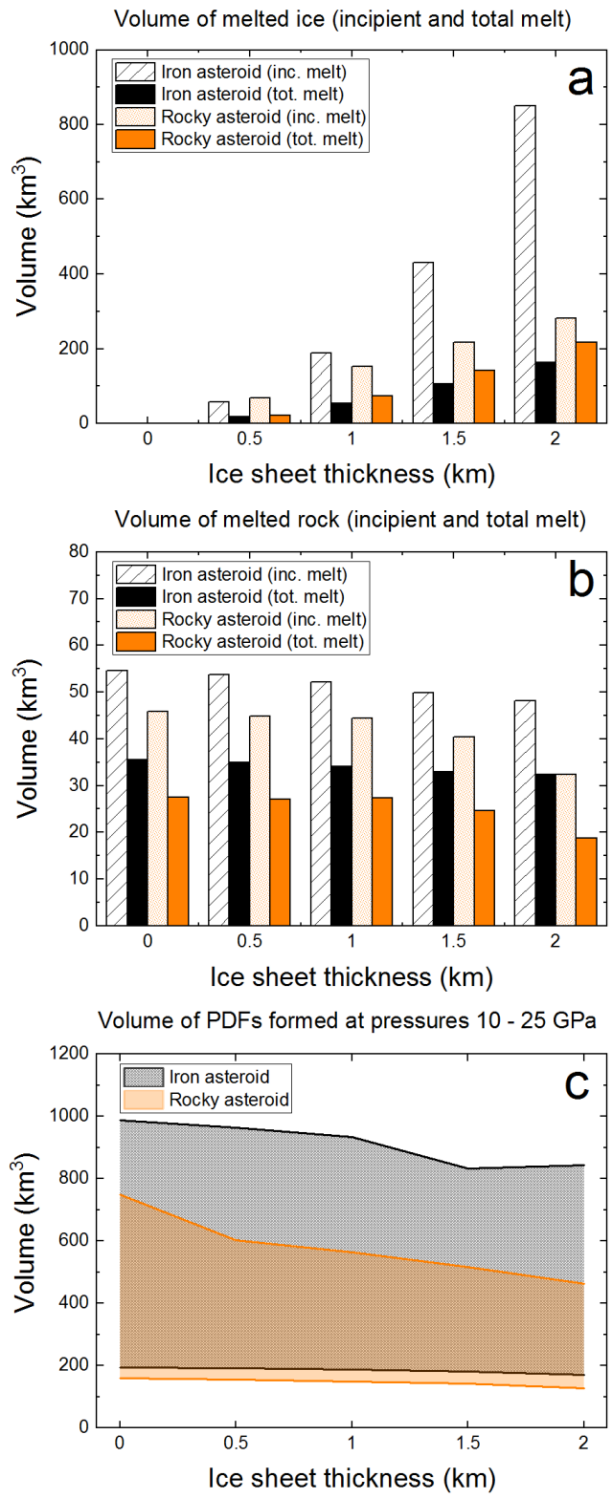

**List of Tables**

**Table 1:** Summary of model parameters for ice sheet, rock (Earth's crust) and iron (impactor). The parameters for ice correspond to Bray et al. (2014), with the exception of the friction coefficient for damaged material, which comes from Bray (2009) fits to laboratory data (Beeman et al., 1988). The parameters for rock are consistent with those listed in Rae et al. (2019). The parameters for the Block Model of acoustic fluidization (Melosh, 1979) correspond to the previous studies (e.g., Collins, 2014; Rae et al., 2019).

| Parameter Description, variable, units | Variable | Units | Ice sheet | Rock | Impactor |
|---|---|---|---|---|---|
| Surface temperature | $T_s$ | K | 250 | 250 | 250 |
| Poisson's ratio | $v$ | - | 0.33 | 0.30 | 0.29 |
| Melt temperature at zero pressure | $T_m$ | - | 273 | 1673 | 1811 |
| Thermal softening coefficient | $\xi$ | - | 1.2 | 1.2 | 1.2 |
| Material constant, Simon $a$ | $a$ | GPa | 1.253 | 6.0 | 6.0 |
| Material constant, Simon $c$ | $c$ | - | 3 | 3 | 3 |
| Cohesion, intact | $Y_{i0}$ | GPa | 0.01 | 0.01 | 0.01 |
| Coefficient of internal friction, intact | $\mu_i$ | - | 2 | 2 | - |
| Limiting strength at high pressure, intact | $Y_{lim}$ | GPa | 0.11 | 2.5 | - |
| Cohesion, damaged | $Y_{d0}$ | MPa | 0.01 | 0.01 | - |
| Coefficient of internal friction, damaged | $\mu_d$ | - | 0.6 | 0.6 | - |
| Limiting strength at high pressure, damaged | $Y_{dlim}$ | GPa | 0.11 | 2.5 | - |
| Acoustic fluidization viscosity constant | $\gamma_\eta$ | - | 0.015 | 0.015 | - |
| Acoustic fluidization time decay constant | $\gamma_\beta$ | - | 300 | 300 | - |
| Equation of state (EOS) | | | ice Tillotson | granit2 ANEOS | iron ANEOS |

**Table 2:** Summary of the results outlining the peak pressures and the impact-induced melt volumes for an iron asteroid and a rocky asteroid. The columns are as follows: [1] ice sheet thickness; shock pressure required for [2] incipient and [3] total melting of ice, and [4] incipient and [5] total vaporization of ice (Pierazzo et al., 1997); [6-7] shock pressure range at which PDFs form (French and Koeberl, 2010); shock pressure required for [8] incipient and [9] total melting of rock (Pierazzo et al., 1997); [10] maximum shock pressure reached within the target rock layer.

| | [1] | [2] | [3] | [4] | [5] | [6] | [7] | [8] | [9] | [10] |
|---|---|---|---|---|---|---|---|---|---|---|
| | Ice sheet thickness [km] | V [km³], P > 0.4 GPa | V [km³], P > 3 GPa | V [km³], P > 4.5 GPa | V [km³], P > 43 GPa | V [km³], P > 10 GPa | V [km³], P > 25 GPa | V [km³], P > 46 GPa | V [km³], P > 56 GPa | Max P [GPa] |
| **Iron Asteroid** | 0 | n/a | n/a | n/a | n/a | 986.8 | 194.3 | 54.6 | 35.5 | 302 |
| | 0.5 | 58.3 | 17.5 | 13.4 | 1.3 | 964.6 | 191.7 | 53.7 | 35.0 | 251 |
| | 1 | 189.2 | 53.9 | 31.6 | 2.2 | 933.6 | 188.1 | 52.1 | 34.1 | 224 |
| | 1.5 | 431.0 | 106.9 | 57.5 | 3.4 | 832.1 | 181.8 | 49.8 | 33.0 | 204 |
| | 2 | 850.7 | 163.8 | 85.4 | 4.7 | 843.2 | 170.6 | 48.1 | 32.3 | 189 |
| **Rocky Asteroid** | 0 | n/a | n/a | n/a | n/a | 749.5 | 159.6 | 45.9 | 27.5 | 222 |
| | 0.5 | 68.4 | 22.5 | 14.6 | 0.6 | 603.3 | 154.7 | 44.8 | 27.0 | 176 |
| | 1 | 153.2 | 74.1 | 43.0 | 1.0 | 563.5 | 149.5 | 44.5 | 27.4 | 157 |
| | 1.5 | 218.0 | 141.4 | 77.8 | 3.9 | 516.6 | 142.0 | 40.4 | 24.7 | 140 |
| | 2 | 282.7 | 217.4 | 119.7 | 9.6 | 462.2 | 127.9 | 32.3 | 18.8 | 126 |